\documentclass[aps,prl,twocolumn,twoside,superscriptaddress,nofootinbib,preprintnumbers,floatfix,longbibliography]{revtex4-2}

\usepackage{hyperref}
\usepackage{graphicx}
\usepackage{amsmath}
\usepackage{mathrsfs}
\usepackage{xspace}
\usepackage{amsfonts}
\usepackage{color}
\usepackage{booktabs,siunitx}

\newcommand{\eq}[1]{Eq.~\eqref{eq:#1}}

\newcommand{\nn}{\nonumber}

\allowdisplaybreaks[2]

\usepackage{xcolor}
\usepackage{marginnote}
\usepackage[normalem]{ulem}
\usepackage{dsfont,mathrsfs}

\definecolor{darkgreen}{rgb}{0.13,0.55,0.13}
\definecolor{or}{rgb}{0.88,0.43,0.02}

\begin{document}
\preprint{FERMILAB-PUB-25-0029-T, MIT-CTP 5827, MCNET-25-01}
\title{QCD Theory meets Information Theory}

\author{Beno\^{i}t Assi}
\email{bassi@fnal.gov}
\affiliation{Theory Division, Fermilab, Batavia, IL 60510, USA}
\affiliation{Department of Physics, University of Cincinnati, Cincinnati, OH, 45221, USA}
\author{Stefan H{\"o}che}
\email{shoeche@fnal.gov}
\affiliation{Theory Division, Fermilab, Batavia, IL 60510, USA}
\author{Kyle Lee}
\email{kylel@mit.edu}
\affiliation{Center for Theoretical Physics -- a Leinweber Institute, Massachusetts Institute of Technology, Cambridge, MA 02139, USA}
\author{Jesse Thaler}
\email{jthaler@mit.edu}
\affiliation{Center for Theoretical Physics -- a Leinweber Institute, Massachusetts Institute of Technology, Cambridge, MA 02139, USA}
\affiliation{The NSF AI Institute for Artificial Intelligence and Fundamental Interactions}

\begin{abstract}
We present a novel technique to incorporate precision calculations from quantum chromodynamics into
fully differential particle-level Monte-Carlo simulations.
By minimizing an information-theoretic quantity subject to constraints, our reweighted Monte Carlo incorporates systematic uncertainties absent in individual Monte Carlo predictions, achieving consistency with the
theory input in precision and its estimated systematic uncertainties.
Our method can be applied to arbitrary observables known from precision calculations, including multiple observables simultaneously.
It generates strictly positive weights, thus offering a clear path to statistically powerful and
theoretically precise computations for current and future collider experiments.
As a proof of concept, we apply our technique to event-shape observables at electron-positron colliders, leveraging existing precision calculations of thrust.
Our analysis highlights the importance of logarithmic moments of event shapes, which have not been previously studied in the collider physics literature.
\end{abstract}
\maketitle

%%%%%%%%%%%%%%%%%%%%%%%%%%%%%%%%%%%%%%%%%%%%%%%%%%%%%%%%%%%%%%%%%%%%%%%%%%%%%%%%

High-energy collider physics relies on Monte-Carlo (MC) simulation programs called event generators, which are the only means to compute observables under arbitrary fiducial cuts and experimental constraints~\cite{Buckley:2011ms}.
Event generators play a crucial role in experimental analyses, in benchmarking new theoretical predictions against experimental data, and in designing future experiments.
On the other hand, their versatility comes at the expense of theoretical precision. Enhancing their formal accuracy and incorporating well-defined systematic uncertainties is  essential for advancing the near- and long-term goals of the field~\cite{EuropeanStrategyforParticlePhysicsPreparatoryGroup:2019qin,Narain:2022qud,Butler:2023glv}. 

%%%%%%%%%%%%%%%%%%%%%%%%%%%%%%%%%%%%%%%%%%%%%%%%%%%%%%%

Analytic calculations for specific collider observables have advanced significantly in parametric accuracy~\cite{Maltoni:2022bqs,Craig:2022cef,Huss:2022ful}, both fixed-order and resummed, involving both quantum chromodynamics (QCD) and electroweak (EW) effects.
These calculations often exceed the precision of event generators, but they are inherently more inclusive, lacking the detailed phase-space dependence necessary for a comprehensive analysis of experimental data.
Consequently, while dedicated calculations offer precise predictions for particular observables, they cannot replace general-purpose MC event generators.

%%%%%%%%%%%%%%%%%%%%%%%%%%%%%%%%%%%%%%%%%%%%%%%%%%%%%%%

To bridge the gap, the community has developed a host of matching and merging algorithms aimed to systematically improve event generators through higher-order computations~\cite{Campbell:2022qmc}.  
However, achieving the same level of precision as dedicated calculations remains a challenge:
these schemes rely on elaborate phase‑space partitions to prevent double counting, often generating sizable fractions of negative‑weight events.
Key obstacles include merging next-to-next-to-leading order (NNLO) calculations at the fully differential level~\cite{
  Hamilton:2012rf,Hamilton:2013fea,Hoche:2014uhw,Karlberg:2014qua,Hamilton:2015nsa,
  Alioli:2015toa,Astill:2018ivh,Re:2018vac,Monni:2019whf,Mazzitelli:2020jio,Alioli:2020qrd,
  Mazzitelli:2021mmm,Alioli:2021qbf,Lindert:2022qdd,Gavardi:2022ixt,Alioli:2023har,Gavardi:2023aco},
and improving parton showers to next-to-leading logarithmic 
(NLL) accuracy~\cite{Hoche:2017iem,Dulat:2018vuy,Dasgupta:2020fwr,Bewick:2019rbu,
  Forshaw:2020wrq,Nagy:2020rmk,Nagy:2020dvz,Dasgupta:2020fwr,Bewick:2021nhc,Gellersen:2021eci,
  vanBeekveld:2022zhl,vanBeekveld:2022ukn,Herren:2022jej,Assi:2023rbu,FerrarioRavasio:2023kyg,
  Preuss:2024vyu,Hoche:2024dee,vanBeekveld:2024wws}
and beyond the leading-color approximation~\cite{Gustafson:1992uh,Giele:2011cb,
  Nagy:2012bt,Platzer:2012np,Nagy:2014mqa,Nagy:2015hwa,Platzer:2018pmd,Isaacson:2018zdi,
  Nagy:2019rwb,Nagy:2019pjp,Forshaw:2019ver,Hoche:2020pxj,DeAngelis:2020rvq,
  Holguin:2020joq,Hamilton:2020rcu}. 
Further, certain theoretical insights --- such as (non)perturbative power corrections~\cite{Moult:2018jjd,Ebert:2018lzn,Dokshitzer:1995zt,Gardi:2000yh,Lee:2006nr,Mateu:2012nk}, nonperturbative (multi-hadron) fragmentation functions~\cite{Metz:2016swz,Chang:2013iba,Lee:2023tkr,Boussarie:2023izj}, and theoretical uncertainties~\cite{Cacciari:2011ze,Tackmann:2024kci,Lim:2024nsk} --- have yet to be fully incorporated into existing MC frameworks.
These multiple challenges call for a more general and scalable strategy to improve MC simulations with precision theoretical knowledge.

%%%%%%%%%%%%%%%%%%%%%%%%%%%%%%%%%%%%%%%%%%%%%%%%%%%%%%%

In this letter, we introduce an information‑theoretic reweighting technique that embeds any event‑level precision calculation into a fully differential MC sample, overcoming the key limitations of traditional matching and merging schemes.  By matching a limited number of theory‑inspired moments, our method yields strictly positive weights, removes the need for phase‑space slicing, and treats multiple observables on equal footing, all within a single numerically efficient algorithm. Moreover, our method propagates the systematic uncertainties of those moments directly into the MC, delivering both enhanced precision and coherent error estimates while preserving full exclusivity of the event simulation.

\emph{\textbf{Information-Theoretic Approach.}}
We denote the MC distribution (or ``prior'') as $q(\Phi)$, where $\Phi$ encapsulates the complete particle-level information, including momenta, charges, and flavors.
Any set of collider observables $\vec{v}=\{v_1,v_2,\ldots\}$ can be computed through $\Phi\to \vec{v}(\Phi)$, particularly those involving realistic experimental constraints and cuts.
Similarly, we denote the desired theoretical distribution for $\vec{v}$ (or ``target'') as $r(\vec{v})$; this distribution can be computed either analytically or numerically, and typically involves an integral over $\Phi$ (or a proxy for it).
The goal of our method is to define an ``ideal'' distribution $p(\Phi)$ that merges the precise theoretical target $r(\vec{v})$ with the fully exclusive MC prior $q(\Phi)$.

In information theory, the Kullback–Leibler (KL) divergence, also known as the relative entropy,
quantifies the difference between two probability distributions.
Specifically, the KL divergence of the ideal distribution $p(\Phi)$ with respect to the prior distribution $q(\Phi)$ is:
\begin{equation}\label{eq:Loss} 
  \mathcal{L}_{\rm KL}(p \parallel q) =
  \int d\Phi \, p(\Phi) \ln \frac{p(\Phi)}{q(\Phi)} \;,
\end{equation}
where the integration measure $d\Phi$ has a complicated phase-space dependence.
Note that $\mathcal{L}_{\rm KL}(p \parallel q) = 0$ implies that the two distributions are identical.

In our approach, the ideal distribution $p(\Phi)$ incorporates theoretical information from the target distribution $r(\vec{v})$ in the form of moments.
Let $g_j(\vec{v})$ be a set of basis functions to be discussed later, and $c_j$ and $d_j$ be their respective expectation values under $r(\vec{v})$ and $p(\Phi)$:
\begin{equation}
\label{eq:c_and_d_defs}
c_j[r] =\int d\vec{v} \, r(\vec{v}) \, g_j(\vec{v})   \;, \quad
d_j[p] =\int d\Phi \, p(\Phi) \, g_j\big(\vec{v}(\Phi)\big) \;.
\end{equation}
To determine $p(\Phi)$, we minimize its KL divergence to the prior $q(\Phi)$, subject to the target constraints $c_j = d_j$ for all $j$.
A similar moment matching protocol was proposed in Ref.~\cite{Desai:2024yft}, which itself was inspired by Boltzmann's approach to statistical mechanics~\cite{e17041971} and Jaynes' maximum entropy principle~\cite{Jaynes:1957zza,Jaynes:1957zz}.

Expressed as a loss, we seek $p(\Phi)$ that extremizes:
\begin{equation}\label{eq:gconst}
\mathcal{L}[p,q] = \mathcal{L}_{\rm KL}(p \parallel q) + \sum_j \lambda_j \big( c_j[r] - d_j[p] \big)\;, 
\end{equation}
where $\lambda_j$ are Lagrange multipliers enforcing the constraints.%
\footnote{
  In practice, we numerically determine the Lagrange multipliers $\lambda_j$ by inserting Eq.~\eqref{eq:pSol} into Eq.~\eqref{eq:c_and_d_defs} and minimizing the loss 
  $\sum_j \left(\frac{c_j-d_j}{c_j+d_j}\right)^2$.
  This balances the relative contributions of all constraints, allowing efficient optimization via ADAM~\cite{kingma2014adam}.}
Assuming the constraints are non-degenerate,%
\footnote{Degeneracies almost never occur in practice, unless one has incompatible constraints on the same observable.}
the distribution $p(\Phi)$
that extremizes $\mathcal{L}$ is~\cite{e17041971}:
\begin{equation}
\label{eq:pSol} 
p(\Phi) = q(\Phi) \, w(\Phi)\;, \quad w(\Phi) \equiv  \exp\bigg[ -\sum_j \lambda_j g_j\big(\vec{v}(\Phi)\big) \bigg]\;. 
\end{equation} 
Since $q(\Phi)$ is fully differential,
and each basis function $g_j$ is defined over all of $\Phi$, the solution $p(\Phi)$ maintains comprehensive phase-space coverage without double counting.
To make practical use of Eq.~\eqref{eq:pSol}, we treat $w(\Phi)$ as event-by-event weights, which are strictly positive 
by construction, preventing the computational bottlenecks associated with negative weights in MC simulations~\cite{Campbell:2022qmc}. 

\emph{\textbf{Logarithmic Moments.}}
Now, we must determine the basis functions $g_j(\vec{v})$ that best characterize the theoretical target distribution $r(\vec{v})$.
To illustrate our approach, we restrict our attention to a single event-shape observable $v$ in electron-positron ($e^+e^-$) collisions, leaving generalizations to future work.
Many event shapes computed to high parametric accuracy in QCD take a log-exponentiated form at leading power (LP) in $v$:%
\begin{align}
\label{eq:LPeq}
r(v)_{\rm LP} &= \frac{d}{dv}\Bigg[\Bigg(1+\sum_{m=1}^{\infty} C_m^{[0]}\left(\frac{\alpha_s}{4 \pi}\right)^m\Bigg) \nonumber\\
&\quad \times\exp \Bigg[\sum_{i=1}^{\infty} \sum_{j=1}^{i+1} G_{i j}\left(\frac{\alpha_s}{4 \pi}\right)^i \ln ^j v\Bigg]\Bigg]\,,
\end{align}
where $\alpha_s$ is the strong coupling constant, and $C_m^{[0]}$ and $G_{ij}$ are $v$-independent coefficients.
The ``Sudakov logarithms'' in the exponent are characteristic of analytic resummation, and they also appear in MC parton shower algorithms.
Comparing Eqs.~\eqref{eq:pSol} and~\eqref{eq:LPeq}, we see that logarithmic moments $g_n(v) = \ln^n v$ form a natural basis to encapsulate information from analytic resummation.%
\footnote{In general, the $G_{ij}$ depend on anomalous dimensions at different perturbative orders~\cite{Almeida:2014uva}.  To better separate these orders, one could use the $v$-dependent functions that multiply the anomalous dimensions.  In practice, we find the logarithmic basis to be more practical and equally efficient for capturing resummation effects.}

To highlight the structure of logarithmic moments, we compute the leading-logarithmic (LL) distribution of the event shape ``thrust''~\cite{Farhi:1977sg,Catani:1992ua} at fixed coupling (f.c.).
The one-minus-thrust ($\tau \equiv 1-T$) distribution is:
\begin{align}
r(\tau)_{\rm LL,f.c.} =  \frac{-2 \alpha_s C_F}{\pi}\frac{\ln \tau}{\tau} \exp \left[-\frac{ \alpha_s C_F}{\pi}\ln ^2 \tau\right]\,,
\end{align}
where $C_F =4/3$ is the quadratic Casimir for quarks in QCD.
This yields the following logarithmic moments:
\begin{align}
\langle \ln^n\tau\rangle_{\rm LL,f.c.} &= \int_0^{\tau_{\rm max}} d\tau\, r(\tau)_{\rm LL,f.c.} \ln^n\tau \nn\\
&=(-1)^n\left(\frac{\pi}{\alpha_s C_F}\right)^{n/2}\Gamma\left[1+\frac{n}{2}\right]\,,
\end{align}
where we take $\tau_{\rm max}=1$ (instead of the physical $\frac{1}{2}$) for simpler illustration.
For generic $n$, $\langle \ln^n\tau \rangle$ features fractional powers of $\alpha_s$, which is characteristic of Sudakov-safe observables~\cite{Larkoski:2013paa,Larkoski:2015lea,Cal:2020flh,Cal:2021fla}.
While logarithmic moments do not have an order-by-order Taylor expansion in $\alpha_s$ (cf.~ordinary moments $\langle \tau^m \rangle$ in e.g.~\cite{Abbate:2012jh}), they are nevertheless well-defined in resummed QCD.
Their sensitivity to resummation effects makes them interesting observables in their own right, which can also be measured and compared with data using e.g.~\cite{Desai:2024yft}.

\begin{figure}[t]
\centering
\includegraphics[width=0.48\textwidth]{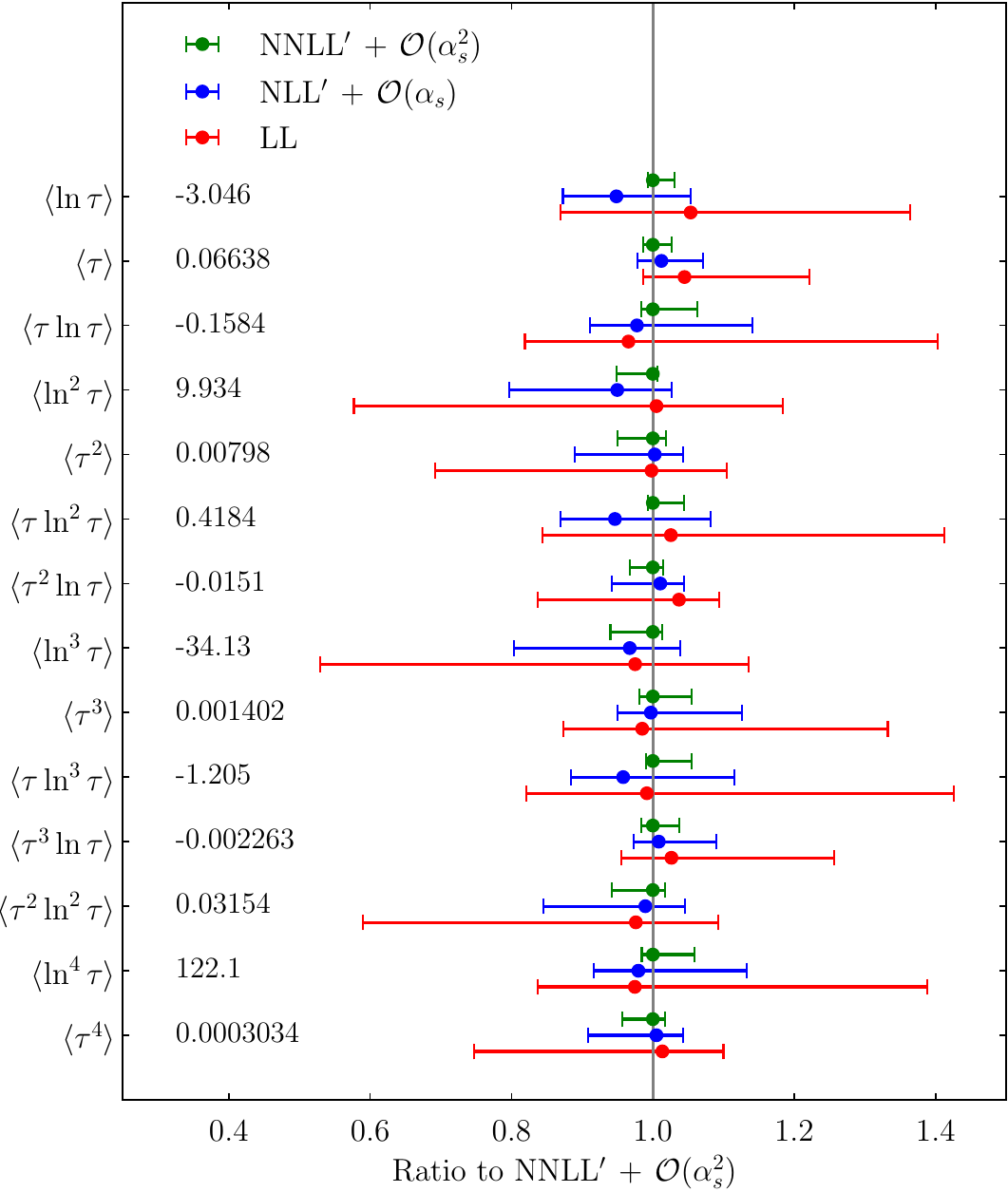}
\caption{Normalized moments $\langle \tau^m \ln^n\tau\rangle$ at LL, NLL$' +\mathcal{O}(\alpha_s)$, and NNLL$' +\mathcal{O}(\alpha_s^2)$ accuracy, with renormalon subtraction and nonperturbative power corrections. The numbers correspond to the central values computed at the highest precision.}
\label{fig:logmom}
\end{figure}

\begin{figure}
\begin{center}
\includegraphics[width=0.48\textwidth]{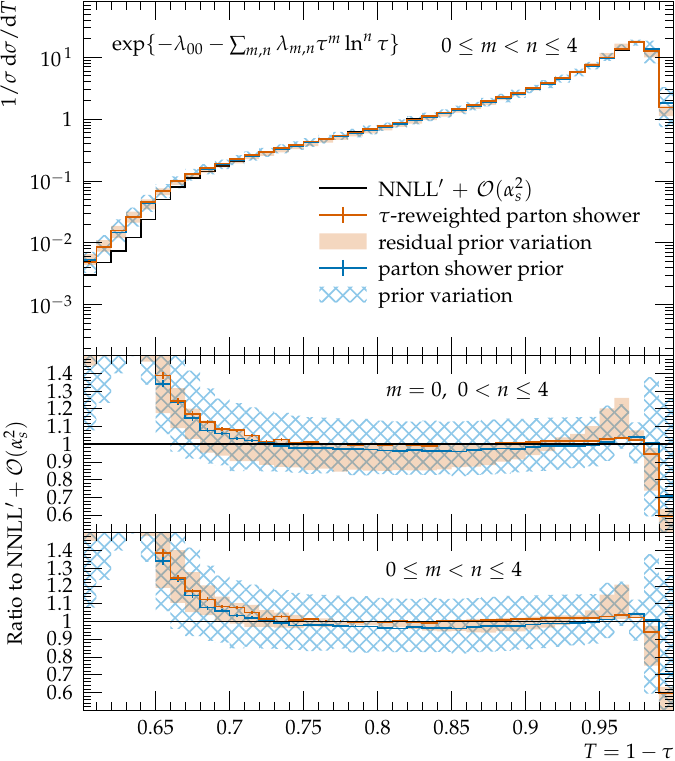}
\end{center}
\caption{
Impact of moment constraints on the MC prior, with the resulting thrust distribution in the top panel.
The middle and bottom panels show the ratio of the reweighted prior to the NNLL+$\mathcal{O}(\alpha_s^2)$ calculation using, respectively, logarithmic moments $\langle \ln^n \tau \rangle$ and mixed moments $\langle \tau^m \ln^n \tau \rangle$.}
\label{fig:priorcollapse}
\end{figure}

\emph{\textbf{Thrust Moments with Uncertainties.}}
Going beyond LP, the order-by-order calculation of a generic Sudakov-log observable $v$ takes the form:
\begin{align}
\label{eq:pdist}
r(v) = \delta(v)&+\sum_{m=1}^{\infty}\sum_{n=1}^{2m-1}k^{\rm LP}_{mn}\left(\frac{\alpha_s}{4\pi}\right)^m \left[\frac{\ln^{n}v}{v}\right]_+\\
&+\cdots+\sum_{m=1}^{\infty}\sum_{n=1}^{2m-1}k^{\rm N^kLP}_{mn}\left(\frac{\alpha_s}{4\pi}\right)^m \frac{\ln^{n}v}{v}v^{k-1}\,,\nonumber
\end{align}
which involves plus-function regularization of the $v \to 0$ singularity.
Resumming just the $k^{\rm LP}_{mn}$ terms recovers Eq.~\eqref{eq:LPeq}.
Taking thrust as an example, this analytic expression motivates choosing basis functions of the form:%
\footnote{Despite the appearance of $v^{-1}$ in  Eq.~\eqref{eq:pdist}, it is preferable to compute moments with $m \geq 0$.  The reason is that the $v^{-1}$ terms generated by $d/dv$ in Eq.~\eqref{eq:LPeq} are well approximated by all but the most pathological MC priors, up to normalization effects.
For fixed-order calculations without resummation, plain moments of $\tau^m$ (i.e.~$n = 0$) are recommended.
}
\begin{align}
\label{eq:momentbasis}
g_{mn}(\tau) = \tau^m\ln^n\tau\;. 
\end{align}

In Fig.~\ref{fig:logmom}, we present several $\langle \tau^m \ln^n\tau\rangle$ thrust moments at NNLL$'$ + $\mathcal{O}(\alpha_s^2)$ accuracy~\cite{Abbate:2010xh,Abbate:2012jh,Bell:2018gce,Benitez:2024nav,Hoang:2014wka}.
Here, we adopt the renormalon subtraction and best-fit values of the nonperturbative power corrections and $\alpha_s(m_Z)$ from Ref.~\cite{Benitez:2024nav}, and estimate uncertainties using their profile functions and scale variations. We observe good convergence of the moments as the perturbative accuracy improves.

\begin{figure*}[t]
\begin{center}
\includegraphics[scale=0.5]{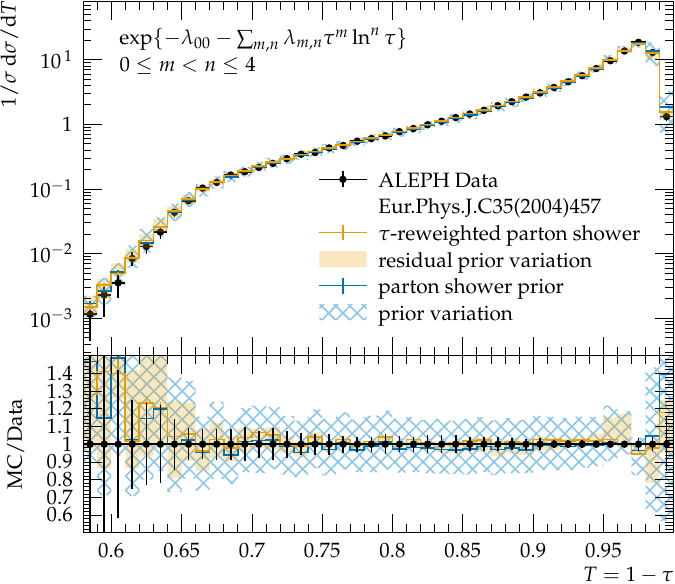}\hfill
\includegraphics[scale=0.5]{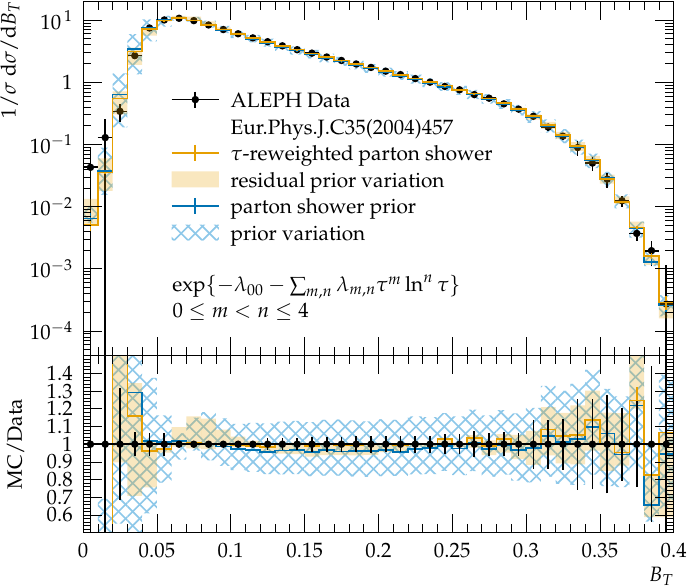}\hfill
\includegraphics[scale=0.5]{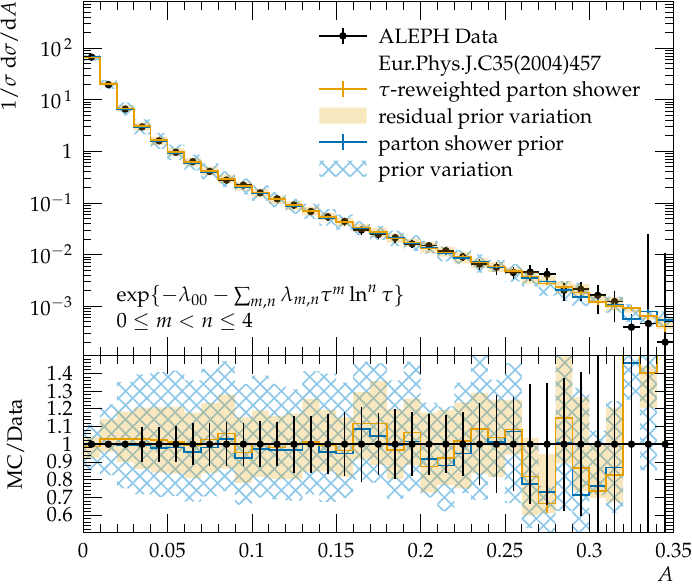}
\end{center}
\caption{Impact of thrust-based reweighting on distributions of (left) thrust, (middle) total jet broadening, and (right) aplanarity, as compared to ALEPH data~\cite{ALEPH:2003obs}.}
\label{fig:simtau}
\end{figure*}

\emph{\textbf{Constraining MC Priors for Thrust.}}
To illustrate the impact of moment-based reweighting, we show how different MC priors converge after augmenting them with precision theoretical targets.
The default prior comes from the Dire parton-shower generator~\cite{Hoche:2015sya}, combined with Lund string fragmentation as implemented in Pythia~\cite{Bierlich:2022pfr}.
Additional priors are created by varying the default values of $\alpha_s(m_Z)=0.118$ up and down by 0.01, and by changing the string hadronization model parameters.
We emphasize that an individual prior does not have any notion of uncertainty, so the envelope of plausible MC variations is one way to define an uncertain band for the MC prior. Importantly, moment reweighting can only ``fix'' priors that already live in the space of consistent distributions --- arbitrarily poor priors cannot be driven to the true distribution by our procedure.
Here and below, we use the event generation framework Sherpa~\cite{Sherpa:2019gpd,Sherpa:2024mfk}
and the analysis framework Rivet~\cite{Buckley:2010ar,Bierlich:2024vqo}.

In Fig.~\ref{fig:priorcollapse}, we show the thrust distribution in $e^+e^-\to \text{hadrons}$ at the $Z$ pole for the range of priors, compared to the NNLL$^{\prime}+\mathcal{O}(\alpha_s^2)$ calculation with the central scale choice.
We see that the prior variations, shown as the blue cross-hatched bands,
deviate significantly from the analytic result.
The orange solid bands show how different moment constraints affect the prior distributions after reweighting.
To guide the eye, the upper panel shows the $T = 1 - \tau$ distributions, but to understand the impact of reweighting, it is more instructive to study ratios to the analytic calculation.

The ratio panels of Fig.~\ref{fig:priorcollapse} show the result of incorporating (middle) logarithmic moments $\langle \ln^n \tau \rangle$ for $0< n \leq 4$ and (bottom) mixed moments $\langle \tau^m \ln^n \tau \rangle$ for $0 \leq m < n \leq 4$.
In both cases, the normalization $\langle 1 \rangle$ is also included as an input, and we do not consider uncertainties on the moments themselves, just the variation from the choice of prior.
With just logarithmic moments, the variation band shrinks considerably near $T \in [0.9, 0.95]$, as expected since they capture resummation effects near the peak region.
These moments are also sensitive to nonperturbative physics as $T \to 1$, which the theory calculation models via a shape function~\cite{Hoang:2014wka, Abbate:2010xh, Hoang:2015hka,Benitez:2024nav,Bell:2018gce}.
To mitigate sensitivity to nonperturbative modeling, we restrict our analysis to $n \leq 4$.
Using mixed moments improves the performance, where the priors exhibit good collapse to the analytic calculations in the peak and tail regions within $T \in [0.7, 0.95]$.
Since the theory calculation is based on dijet factorization, we limit our analysis to $m < n$, since the very far-tail region with $T \in [\frac{1}{2}, \frac{2}{3}]$ is affected by unmodeled physics. In the End Matter, we describe in detail the theoretical ingredients necessary to apply our method to the full phase space.

\emph{\textbf{Impact of Reweighting on Other Observables.}}
The result of our approach is a weight function $w(\Phi)$, which can be applied to any MC distribution, not just those related to the target distribution $r(\vec{v})$.
As an example, it is instructive to see how weights derived from thrust moments affect other observables, particularly those with different behaviors under resummation.
In  Fig.~\ref{fig:simtau}, we show the impact of moment reweighting on thrust ($T=1-\tau$) \cite{Farhi:1977sg}, total jet broadening ($B_T$) \cite{Rakow:1981qn}, and aplanarity ($A$) \cite{Bjorken:1969wi}, as compared to data from the ALEPH experiment~\cite{ALEPH:2003obs}.
Here, we use the same mixed moment constraints on $\langle \tau^m \ln^n \tau \rangle$ as in Fig.~\ref{fig:priorcollapse}, using the central scale choice at NNLL$^{\prime}+\mathcal{O}(\alpha_s^2)$.

For all three observables, the MC prior variation is consistent with ALEPH data up to experimental uncertainties.
As expected, thrust-based reweighting shrinks the residual variation for the thrust distribution.
Remarkably, similar improvements are seen for broadening; even though thrust and broadening have different anomalous dimensions,  thrust moments are apparently sufficient to constrain the dijet phase space probed by broadening. 
By contrast, thrust reweighting does not significantly affect the aplanarity distribution, as expected since aplanarity is a genuinely multijet observable.
Of course, simultaneously precise predictions for thrust and aplanarity could be obtained by including moments like $\langle A^m \ln^n A \rangle$ in Eq.~\eqref{eq:pSol}, as well as mixed $\tau$--$A$ moments. As long as the chosen moment constraints remain compatible, our method generalizes to multiple observables with no extra complexity.
This highlights a particular strength of our method, which overcomes the limitations of single-observable-based approaches to NNLO matching~\cite{Alioli:2012fc,Hamilton:2013fea,Monni:2019whf}.

\begin{figure}[t!]
\begin{center}
\includegraphics[width=0.48\textwidth]{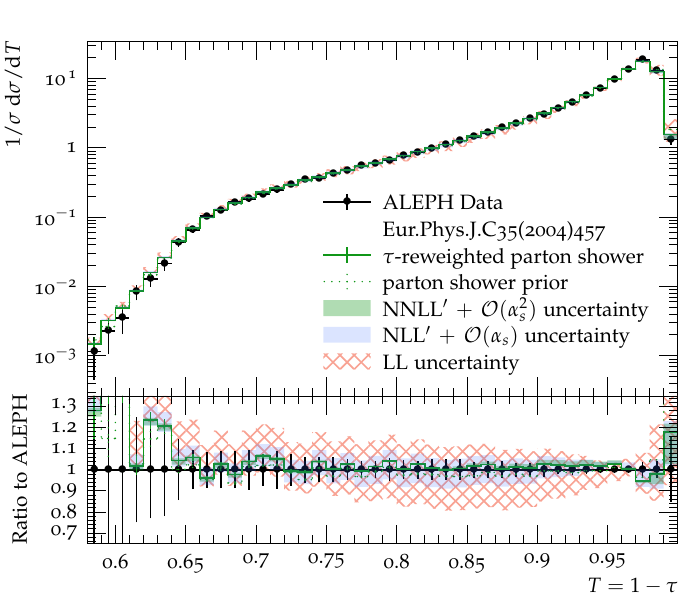}
\end{center}
\caption{Impact of scale variations on the NNLL$'+\mathcal{O}(\alpha_s^2)$ moments propagated to the event weights via the $\lambda_i$ in Eq.~\eqref{eq:pSol}. The prior distribution shown is Dire~\cite{Hoche:2015sya} with Pythia~8 hadronization~\cite{Bierlich:2022pfr}. Lower panel: ratio to ALEPH data~\cite{ALEPH:2003obs} for this prior. A comparison across multiple priors is provided in the End Matter.
}
\label{fig:uncertainties}
\end{figure} 

\emph{\textbf{Incorporation of Theoretical Uncertainties.}}
The proper characterization of theory uncertainties in MC generators is an active and important topic of research.
One clear advantage of any approach based on a finite set of moments is that correlated uncertainties can be derived from a finite-dimensional covariance matrix.
Once these (correlated) moment uncertainties have been defined and computed, there is a separate question of how to incorporate them into MC generators.
Our approach provides a natural way to do so, since it is straightforward to propagate uncertainties on the moments to uncertainties on the Lagrange multipliers in Eq.~\eqref{eq:pSol}.

As a proof of concept, we start from the moment uncertainties in Fig.~\ref{fig:logmom}, which just include perturbative scale variations. Each scale variation choice yields a set of single-valued moments, and the resulting values are therefore correlated across variations. For simplicity, however, we present them in the figure as an envelope representing the spread across individual variations, which does not explicitly reflect these correlations.
For the main demonstration, we use a single representative prior:
Dire~\cite{Hoche:2015sya} with Pythia~8 hadronization~\cite{Bierlich:2022pfr},
configured with a two-loop running coupling with $\alpha_s(m_Z)=0.118$ and the
CMW scheme~\cite{Catani:1990rr}. A detailed comparison across four distinct
priors (CSShower~\cite{Schumann:2007mg} vs.\ Dire, combined with Pythia~8
vs.\ Ahadic~\cite{Chahal:2022rid}) is provided in the End Matter.
Note, however, that even these best-in-class priors are formally only LL accurate.

In Fig.~\ref{fig:uncertainties}, we show the reweighted distribution for thrust, with uncertainty bands from propagating scale variation of the moments to the event weights.
We use the same mixed moment constraints on $\langle \tau^m \ln^n \tau \rangle$ as in Fig.~\ref{fig:priorcollapse}.
The uncertainties here are significantly reduced compared to Fig.~\ref{fig:priorcollapse}, though, since the computed uncertainties at NNLL$^{\prime}+\mathcal{O}(\alpha_s^2)$ in Fig.~\ref{fig:logmom} are smaller than naive prior variation.
We reiterate that our $g_{mn}$ basis with $0\leq m<n\leq 4$ is not particularly sensitive to $\tau \simeq 0$ ($T \simeq 1$) and $\tau \gtrsim \frac{1}{3}$ ($T \lesssim \frac{2}{3}$).
Thus, these regions of phase space are dominantly modeled by the prior, and the uncertainties are therefore underestimated.
Away from these extremes, though, we see the expected convergence of the distribution in going from LL, to NLL$^{\prime}+\mathcal{O}(\alpha_s)$, to NNLL$^{\prime}+\mathcal{O}(\alpha_s^2)$ accuracy.
These results could be systematically improved by taking higher $m$ or $n$ moments, or by including moments of additional observables.

\emph{\textbf{Outlook.}}
In this letter, we introduced an information-theoretic framework that systematically embeds theoretical constraints --- obtained from high-precision analytic calculations ---
into general-purpose MC event generators.
Our work addresses the longstanding challenge of
bridging the gap in theoretical precision analytic calculations and event generators, helping to foster a synergistic relationship between these two research domains.

There are multiple promising avenues to explore next.
First, one can apply this approach to state-of-the-art MC generators~\cite{Campbell:2022qmc}
and the highest-precision analytic calculations~\cite{Maltoni:2022bqs,Craig:2022cef,Huss:2022ful},
pushing the accuracy frontier in realistic collider analyses.
This includes precision calculations of inclusive observables, such as energy-energy correlators~\cite{Basham:1978bw,Basham:1978zq} and inclusive jet spectra~\cite{Aversa:1988fv,Ellis:1988hv}, which can be defined at the event level by taking moments.
Second, our method is readily extensible to more complicated multi-differential distributions or new classes of observables, including those that are difficult to calculate analytically but where theoretical insights still exist in partial or factorized forms.
Third, uncertainties arising from scale choices, higher-order corrections, and nonperturbative modeling can be more rigorously incorporated and correlated, providing a clearer understanding of the theoretical error budget.
Finally, the flexibility of the event generator remains intact, making it straightforward to apply these improved predictions to complex experimental analyses, stringent tests of the Standard Model, and searches for physics beyond it.

\section*{Acknowledgments}

We thank Lance Dixon, Andrew Larkoski, Christopher Lee, Ian Moult, Iain Stewart, and Xiaoyuan Zhang for useful discussions, Krish Desai and Benjamin Nachman for information-theoretic insights, and the anonymous referees for valuable feedback.
K.L.\ was supported by the U.S. Department of Energy, 
Office of Science, Office of Nuclear Physics from DE-SC0011090.
J.T.\ was supported by the National Science Foundation under Cooperative Agreement
PHY-2019786 (The NSF AI Institute for Artificial Intelligence and Fundamental Interactions, 
\url{http://iaifi.org/}), by the U.S. Department of Energy Office of High Energy Physics under grant
number DE-SC0012567, and by the Simons Foundation through Investigator grant 929241.
This manuscript has been authored by Fermi Forward Discovery Group, LLC under Contract No. 89243024CSC000002 with the U.S. Department of Energy, Office of Science, Office of High Energy Physics.
This research used resources of the National Energy Research Scientific Computing Center (NERSC), 
a Department of Energy Office of Science User Facility using NERSC award ERCAP0028985.
This work was performed in part at the Aspen Center for Physics, with support for B.A.\ by a grant from the Simons Foundation (1161654, Troyer).
This research was supported in part by grant NSF PHY-2309135 to the Kavli Institute for Theoretical Physics (KITP).

\begin{figure*}[t!]
\begin{center}
\includegraphics[width=0.48\textwidth]{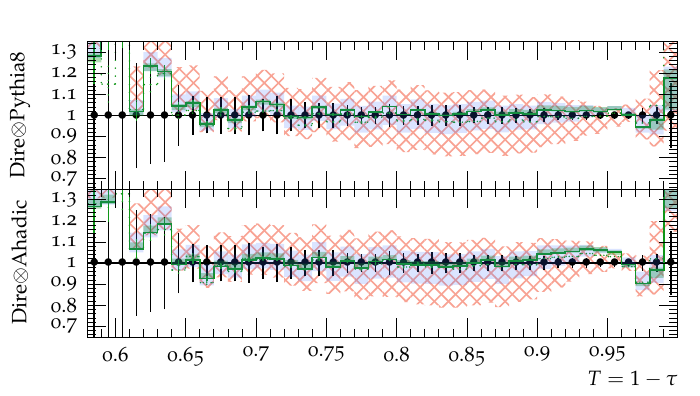}
\includegraphics[width=0.48\textwidth]{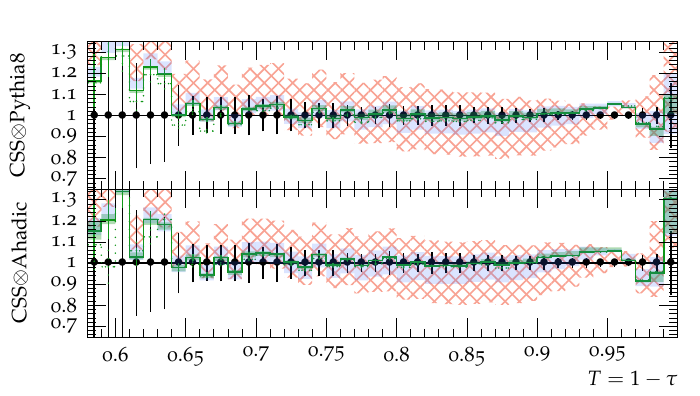}
\end{center}
\caption{Impact of NNLL$'+\mathcal{O}(\alpha_s^2)$ moment scale variations
propagated to the event weights via Eq.~\eqref{eq:pSol}, shown for four priors---
CSShower~\cite{Schumann:2007mg} and Dire~\cite{Hoche:2015sya} combined with
Pythia~8~\cite{Bierlich:2022pfr} or Ahadic~\cite{Chahal:2022rid}---as ratios to ALEPH
data~\cite{ALEPH:2003obs}. Bands reflect only moment-level scale
variations (with correlations preserved) and do not include prior-variation
envelopes.}
\label{fig:uncertainties_2}
\end{figure*}

\section*{End Matter}
\label{endmatter}
\emph{\textbf{Going beyond the region $0.05 \lesssim \tau \lesssim 1/3$.}} In the main text, we use mixed moments $\langle \tau^m \ln^n \tau \rangle$ with $0\le m<n\le 4$ to reweight the thrust distribution. This basis is defined on $\tau\in[0,\tfrac12]$ and provides controlled emphasis across phase space: increasing $m$ shifts weight toward the large-$\tau$ tail, while increasing $n$ enhances sensitivity near the Sudakov peak and into the small-$\tau$ region. The restriction $m<n$ limits far-tail dominance, and the truncation at $n\le 4$ avoids over-weighting the deep endpoint. This choice is therefore tailored to target the window $0.05\lesssim\tau\lesssim\tfrac13$ where dijet factorization is under perturbative control~\cite{Benitez:2024nav,Abbate:2010xh}.

While our moment basis targets $0.05\!\lesssim\!\tau\!\lesssim\!\tfrac13$, our framework applies to the full $\tau$ range, provided that the necessary theory inputs exist. Outside this window, theoretical understanding of the $\tau$ distribution remains under active development. For $\tau\!\lesssim\!0.05$, a standardized thrust shape-function treatment (choice of functional basis and renormalon-subtraction scheme) with fitted parameters and their covariance is required and is currently being explored. For $\tau\!\gtrsim\!\tfrac13$, resummation of Sudakov-shoulder logarithms with validated matching to the three-jet region and documented three-jet power corrections, including a prescription for correlated profile-scale variations is needed~\cite{Catani:1997xc,Bhattacharya:2022dtm,Nason:2023asn}. As these ingredients mature, one can simply augment the basis with additional moments, e.g.\ higher linear or logarithmic moments, to constrain both regions. In the deeply nonperturbative regime, where no analytical formula like~\eq{pdist} exists, it is natural to ask which moments best capture the region.  For example, expanding the shape function in Legendre polynomials~\cite{Ligeti:2008ac,Abbate:2010xh,Bell:2018gce} is a common choice and directly suggests specific moment forms. It is important for the community to explore what alternative function bases are best suited to such regions. Attaining these predictions with full uncertainty correlation matrices would enable extending our constraints across $\tau\!\in[0,\tfrac12]$ without altering the core reweighting algorithm.

Alternatively, we could have assigned large uncertainties outside of the $0.05\!\lesssim\!\tau\!\lesssim\!\tfrac13$ window to reflect current theoretical limitations and propagate them via extra moments in the MC. Concretely, this would mean adding moments with support in $\tau\!\lesssim\!0.05$ and $\tau\!\gtrsim\!\tfrac13$, attaching conservative, correlated variations to the corresponding theory inputs (such as the shape-function and shoulder-resummation settings used in the analytic calculation), and carrying their covariance through the same weight determination as in the main text, yielding much  wider uncertainty bands in those regions. In this work, we decided not to include such additional moments; predictions outside of the central window are therefore prior-dominated and shown just for completeness. We emphasize that none of these choices alters the underlying method; instead they simply reflect the current (lack of) precision for theoretical thrust inputs.

\emph{\textbf{Prior variation versus data.}} 
In the main text, uncertainty
propagation was shown for a single representative prior in Fig.~\ref{fig:uncertainties}. Here we repeat the
procedure for four priors---CSShower~\cite{Schumann:2007mg} and
Dire~\cite{Hoche:2015sya}, each combined with Pythia~8~\cite{Bierlich:2022pfr}
or Ahadic~\cite{Chahal:2022rid}. All priors use two-loop running with
$\alpha_s(m_Z)=0.118$ and the CMW scheme~\cite{Catani:1990rr}. 
Figure~\ref{fig:uncertainties_2} shows, for each prior compared to ALEPH
data~\cite{ALEPH:2003obs}, uncertainty bands obtained solely
from the correlated scale variations of the NNLL$'+\mathcal{O}(\alpha_s^2)$
moments propagated to the Lagrange multipliers in Eq.~\eqref{eq:pSol}; no
prior-variation envelope is included. The residual spread between priors after reweighting
thus quantifies remaining prior dependence. To emphasize just the uncertainties arising from the moments, we consider one prior at a time, though in practice, one would likely incorporate prior variation into the total reported theoretical uncertainty.

%%%%%%%%%%%%%%%%%%%%%%%%%%%%%%%%%%%%%%%%%%%%%%%%%%%%%%%%%%%%%%%%%%%%%%%%%%%%
\bibliography{biblio.bib}{}

%%%%%%%%%%%%%%%%%%%%%%%%%%%%%%%%%%%%%%%%%%%%%%%%%%%%%%%%%%%%%%%%%%%%%%%%%%%%%%%%

% shift all counters by 1000 for uniqueness
\setcounter{equation}{1000}
\setcounter{figure}{1000}
\setcounter{table}{1000}

% change formatting of counters to have an "S" for supplement in front
\renewcommand{\theequation}{S\the\numexpr\value{equation}-1000\relax}
\renewcommand{\thefigure}{S\the\numexpr\value{figure}-1000\relax}
\renewcommand{\thetable}{S\the\numexpr\value{table}-1000\relax}

\end{document}